# Nonreciprocity and One-Way Topological Transitions in Hyperbolic Metamaterials

Alex Leviyev[1*], Binyamin Stein[1*], Tal Galfsky[1,2,3], Harish Krishnamoorthy[1,3], Igor L. Kuskovsky[1,3], Vinod Menon[1,2,3], Alexander B. Khanikaev[1,2#]

E-mail: [#]akhanikaev@qc.cuny.edu

*These authors equally contributed to the present work

[1]Department of Physics, Queens College of The City University of New York, Queens, New York 11367, USA
[2]Department of Physics, City College of The City University of New York, New York, NY 10031, USA
[3]The Graduate Center of The City University of New York, New York, New York 10016, USA

**Control of the electromagnetic waves in nano-scale structured materials is central to the development of next generation photonic circuits and devices. In this context, hyperbolic metamaterials, where elliptical isofrequency surfaces are morphed into surfaces with exotic hyperbolic topologies when the structure parameters are tuned, have shown unprecedented control over light propagation and interaction. Here we show that such topological transitions can be even more unusual when the hyperbolic metamaterial is endowed with nonreciprocity. Judicious design of metamaterials with reduced spatial symmetries, together with the removal of time-reversal symmetry through magnetization, is shown to result in nonreciprocal dispersion and one-way topological phase transitions in hyperbolic metamaterials.**

Optical metamaterials, artificial media with engineered electromagnetic response realized through the structuring on the subwavelength scale, allow achieving properties normally limited or not found in naturally occurring materials [1-6]. One such property is optical nonreciprocity, a rare and generally weak characteristic of light to differentiate between opposite propagation directions [7,8]. This property, which exists in magnetic materials such as ferrites, is of immense importance for devices such as optical isolators and circulators, widely used to stabilize laser operations and to route signals in optical telecommunication networks [9,10,11]. Because of their importance for applications, nonreciprocal optical components are of significant interest for optical integration which can be achieved by combining magneto-optical materials with resonant nanophotonic [12-17] and plasmonic [18-26] elements. Such integration brings significant benefits allowing enhancement of generally weak magneto-optical response of ferrites through strong light-matter interactions in plasmonic and photonic nanostructures, and strongly nonreciprocal response has been demonstrated for a variety of systems from magnetic photonic crystals [12-17] to plasmonic nanostructures and [18-26] metamaterials [27].

Metamaterials with magnetic sub-constituents have been used to achieve negative index of refraction and tunable electromagnetic response of metamaterials in external magnetic field [28-33]. However, very little is known on the nonreciprocal effects that can be engineered using such magnetic metamaterials [27]. In this context hyperbolic metamaterials [34-44,31], a class of metamaterials with hyperbolic isofrequency contours, can be exceptional candidates offering both enhancements of nonreciprocal effects and broadband operation, unattainable in other classes of metamaterials. We demonstrate here that hyperbolic metamaterials with magneto-optical activity exhibit unprecedented nonreciprocal characteristics such as one-way topological transitions [42] and one-way hyperbolic dispersion regimes.

Optical nonreciprocity is a subtle phenomenon, which occurs only when the optical system lacks both time-reversal symmetry and the inversion symmetry [13,15,16]. In the particular case of layered media magnetized in the Voigt geometry, such as photonic crystals or hyperbolic metamaterials studied here (shown in Fig. 1a), the nonreciprocity can be achieved for p-polarized light either through inhomogeneous magnetization [15] or multilayered configuration [16]. Here we consider an archetype system consisting of a three layered metamaterial, structure which is also easy to implement experimentally. The metamaterial is formed by periodically stacking unit cells consisting of a plasmonic ($\epsilon_2 < 0$) layer sandwiched in between two dielectric ($\epsilon_1 \neq \epsilon_3, \epsilon_1 > 0, \epsilon_3 > 0$) layers. We assume that the structure is subject to DC magnetic field $\boldsymbol{B}$ along the $y$-direction (the Voigt geometry). In this case the plasmonic material can be described by a dielectric permittivity tensor of the form $\hat{\epsilon}_2 = [\epsilon_2(\omega), 0, i\Delta_2(\omega); 0, \epsilon_{2,yy}(\omega), 0; -i\Delta_2(\omega), 0, \epsilon_2(\omega)]$ [20], where $\epsilon_2 = \epsilon_\infty - \frac{\omega_p^2}{(\omega + i\gamma)^2 - \omega_B^2} \times \left(1 + i\frac{\gamma}{\omega}\right)$, $\Delta_2 = i\frac{\omega_B}{\omega} \times \frac{\omega_p^2}{(\omega + i\gamma)^2 - \omega_B^2}$, $\epsilon_{2,yy} = \epsilon_\infty - \frac{\omega_p^2}{\omega(\omega + i\gamma)}$, $\epsilon_\infty$ is the high-frequency permittivity, $\omega_p$ is the bulk plasma frequency, $\gamma$ is the decay frequency, $\omega_B = \frac{e}{m^*} B$ is the cyclotron frequency, and $e$ and $m^*$ are the charge and the effective mass of the electron, respectively. It is worth noting that achieving strong nonreciprocity (implying large values of $\Delta_2$) in metals requires strong DC magnetic fields on the order of 1T and stronger [20]. However, it is worth mentioning that the nonreciprocal regimes reported here also occur in hyperbolic metamaterials made of highly doped semiconductors [35] in terahertz and infrared spectral domains for lower values of applied DC magnetic fields, as well as in optical hyperbolic metamaterials containing magneto-optical dielectric components, such as Yttrium Iron Garnet-based ferrites [12-20], in magnetic fields sufficient to provide saturation in the ferrite magnetization.

As the first step, starting with the exact transfer matrix technique, we develop an analytic effective medium theory with the nonreciprocal corrections induced by the magnetization (Supplement A). For reciprocal structures ($\Delta_2 = 0$), this procedure results in the well-known expression $K_z^2/\epsilon_\parallel + k_x^2/\epsilon_\perp = k_0^2$, where $\epsilon_\parallel = \epsilon_1 f_1 + \epsilon_2 f_2 + \epsilon_3 f_3$ and $\epsilon_\perp = (f_1/\epsilon_1 + f_2/\epsilon_2 + f_3/\epsilon_3)^{-1}$ are the effective permittivities parallel and perpendicular to the layers, respectively, and $f_m = a_m/a_0$ is the volume fraction of the $m$-th layer. Similar procedure for the magnetized structure gives rise to two additional terms that are of odd order in the wavenumbers, and the resultant effective medium expression assumes the form:

$$\frac{K_z^2}{\epsilon_\parallel} + \frac{k_x^2}{\epsilon_\perp} + k_x^3 a_0 \frac{\Delta_2}{\epsilon_2} \frac{f_1 f_2 f_3}{\epsilon_\parallel} \left(\frac{\epsilon_3}{\epsilon_1} - \frac{\epsilon_1}{\epsilon_3}\right) = k_0^2 \left[1 + k_x a_0 \frac{f_1 f_2 f_3}{\epsilon_\parallel} \frac{\Delta_2}{\epsilon_2} (\epsilon_3 - \epsilon_1)\right]. \qquad (1)$$

It is worth mentioning that the terms of the odd order in $k_x$ do not appear in the reciprocal case and the next term would be of the fourth order, which is the direct consequence of the reciprocity $k_0(\mathbf{k}) = k_0(-\mathbf{k})$ [or $\omega(\mathbf{k}) = \omega(-\mathbf{k})$]. The latter condition is clearly not satisfied for Eq. (1) where the two addition terms, linear and cubic with respect to $k_x$, lead to the nonreciprocal dispersion $k_0(\mathbf{k}) \neq k_0(-\mathbf{k})$. It can also be seen from Eq.(1) that these terms (and the resultant nonreciprocity) increase with the magneto-optical parameter $\Delta_2$ and/or the dielectric contrast between two layers adjacent to the magneto-plasmonic layer. The latter dependence also confirms that for the inversion symmetric and bilayer structures ($\epsilon_3 = \epsilon_1$) no nonreciprocal response is expected and higher dielectric contrast is desirable for stronger nonreciprocity.

The first and the third orders of the nonreciprocal contributions to Eq.(1) suggest that each will dominate in different domains of values of $k_x$. Thus, for small values of $k_x \ll K_z$, i.e. at the near normal incidence, the linear term in $k_x$ on the right hand side will dominate. It's contribution is rather trivial and can be understood as a horizontal shift of the dispersion curves (either in elliptical or hyperbolic regimes) since $k_x^2 - \alpha k_x \approx (k_x - \alpha/2)^2$, where $\alpha = \frac{\Delta_2}{\epsilon_2} \frac{a_1 a_2 a_3}{a_0} \frac{\epsilon_\perp}{\epsilon_\parallel}(\epsilon_3 - \epsilon_1)$ is a small parameter. For large values of $k_x \gg K_z$, which corresponds to the case of large wavenumbers and is of primary interest in the hyperbolic regime, the cubic term is expected to dominate instead.

The effective medium of layered hyperbolic metamaterials is known to be very good approximation for the case of subwavelength-thick layers considered here. However, this approximation can be further improved by considering higher order terms in the expansions leading to Eq.(1) (Supplement A) [45]. In this case one obtains the effective medium which in general can be recast to the same effective medium expression $K_z^2/\tilde{\epsilon}_\parallel + k_x^2/\tilde{\epsilon}_\perp = k_0^2$, where the effective medium permittivities become functions of the wave-vectors $\tilde{\epsilon}_{\parallel,\perp} = \tilde{\epsilon}_{\parallel,\perp}(K_z, k_x)$, i.e. the medium exhibits nonlocal dielectric response. It is instructive to mention that the first and third order terms of Eq.(1) leading to the nonreciprocity can also be described as magnetization induced nonlocality with the $k_x$-dependent effective parameters

$$\tilde{\epsilon}_\parallel(k_x) = \epsilon_\parallel \left[1 + k_x \frac{a_0 f_1 f_2 f_3}{\epsilon_\parallel} \frac{\Delta_2}{\epsilon_2}(\epsilon_3 - \epsilon_1)\right], \quad (2a)$$

$$\tilde{\epsilon}_\perp(k_x) = \epsilon_\perp \left[\frac{1 + k_x a_0 \frac{f_1 f_2 f_3}{\epsilon_\parallel} \frac{\Delta_2}{\epsilon_2}(\epsilon_3 - \epsilon_1)}{1 + k_x a_0 f_1 f_2 f_3 \frac{\Delta_2 \epsilon_\perp}{\epsilon_2 \epsilon_\parallel}\left(\frac{\epsilon_3}{\epsilon_1} - \frac{\epsilon_1}{\epsilon_3}\right)}\right]. \quad (2b)$$

The difference between this magnetization induced *nonreciprocal nonlocality* and one stemming from higher order expansion terms of reciprocal hyperbolic media is that the effective parameters are odd functions of the wave-vector components as opposed to even functions in the reciprocal nonlocal theories.

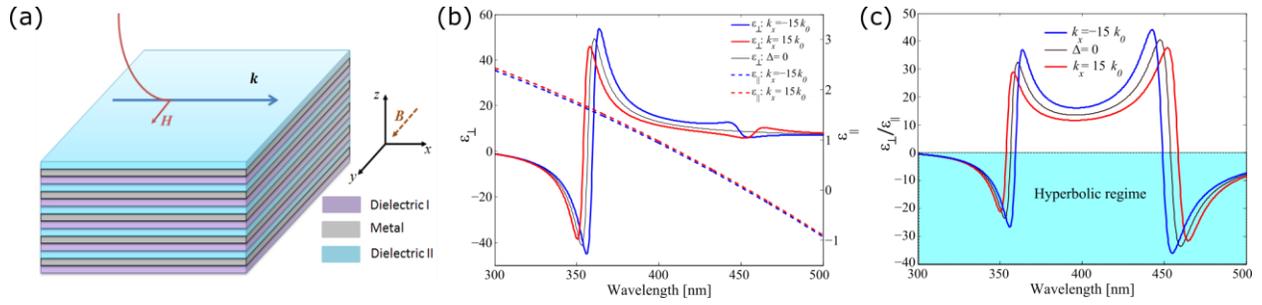

Figure 1| (a) Schematic view of three-layer nonreciprocal metamaterial and the magnetization geometry used (shown by dashed arrow). (b)-(c) Effective permittivities calculated from Eqs.2(a,b) for the three-layer $TiO_2/Ag/SiO_2$ nonreciprocal hyperbolic metamaterial. Blue and red lines correspond to forward ($k_x > 0$) and backward ($k_x < 0$) propagation, respectively, while black lines correspond to the case of nonmagnetic (reciprocal) effective medium theory. Structure parameters are: 14nm-thick $TiO_2$ layer with $n_1 = 2.56$, 20 nm silver layer with Drude parameters $\epsilon_\infty = 4.09$, the plasma frequency $\omega_p = 1.33 \times 10^{16}$ [rad/s], and the damping frequency $\gamma = 1.13 \times 10^{14}$ [rad/s], and magneto-optical parameter $\Delta = 0.1\epsilon_2$, and 14 nm-thick $SiO_2$ layer with $n_3 = 1.46$.

Figure 1 shows the effect of nonreciprocal corrections to the effective medium parameters $\tilde{\epsilon}_\parallel(k_x)$ and $\tilde{\epsilon}_\perp(k_x)$ found from Eqs.2(a,b) as the function of wavelength for two opposite propagation directions with $k_x = 15 \times k_0$ and $k_x = -15 \times k_0$, respectively, for a metamaterial with the unit cell consist of the silver film sandwiched in between lower index quartz and higher index titanium dioxide layers. As can be seen from the dashed lines of Fig.1(b), the effect of magneto-optical activity on $\tilde{\epsilon}_\parallel(k_x)$ is marginal and the red and blue dashed curves corresponding to the magnetic case closely follow each other indicating rather weak nonreciprocity. On the other hand, the effect of nonreciprocal corrections to $\tilde{\epsilon}_\perp(k_x)$ is quite significant. As can be seen from Fig. 1(b), $\tilde{\epsilon}_\perp(k_x)$ has a pole near the wavelength $\lambda$=360 nm, and the shape of the curve near this pole strongly depends on the sign of $k_x$, i.e. on the propagation direction. Another peculiarity occurs near $\lambda$=450 nm, where $\tilde{\epsilon}_\perp$ exhibits additional variations which are also direction dependent. These variations are related to the *epsilon-near-zero* ($\epsilon_\parallel \approx 0$) condition and originate in the presence of $\epsilon_\parallel$ in the nonreciprocal terms in Eqs.(1) and (2). One of the most interesting consequences of nonreciprocity is that near such a pole the metamaterial can exhibit effective permittivities of the opposite sign for the two opposite propagation directions, i.e. $\tilde{\epsilon}_\perp > 0$ for the forward propagation $k_x > 0$, and $\tilde{\epsilon}_\perp < 0$ for the backward propagation $k_x < 0$, or vice versa, as illustrated by Fig.1(c). It can be seen for this plot that two hyperbolic regimes which occur in the structure, Type-II regime at longer wavelengths and Type-I regime for shorter wavelengths, respectively, are spectrally shifted for opposite directions of propagation. And as a result, the onsets of these hyperbolic regimes for forward and backward propagating waves take place at different wavelengths.

For illustrative purposes we will first examine the metamaterial at the wavelength $\lambda = 455$ nm, which in the absence of the external magnetization exhibits the elliptical regime $\{\tilde{\epsilon}_\parallel > 0, \tilde{\epsilon}_\perp > 0\}$, and study how the magnetization affects the dispersion calculated with the use of the nonreciprocal effective medium theory Eq.(1). Figure 2 shows changes in the dispersion as the off-diagonal component of the metal's permittivity $\Delta_2$ gradually increases, which is equivalent to an increase in the applied DC magnetic field. It is seen from Fig.2(a) that as the value of $\Delta_2$ increases, the isofrequency contours acquire progressively more asymmetric shapes, ultimately leading to *a complete change in their topology*. This magnetization induced *topological transition* is more clearly revealed in Fig.2(b) which shows how one side of the closed elliptical contour ($k_x < 0$) gradually changes to the open Type-II hyperbolic contour $\{\tilde{\epsilon}_\| < 0, \tilde{\epsilon}_\perp > 0\}$, while the opposite side ($k_x > 0$) of the contour remains elliptical and only slightly changes in the radius.

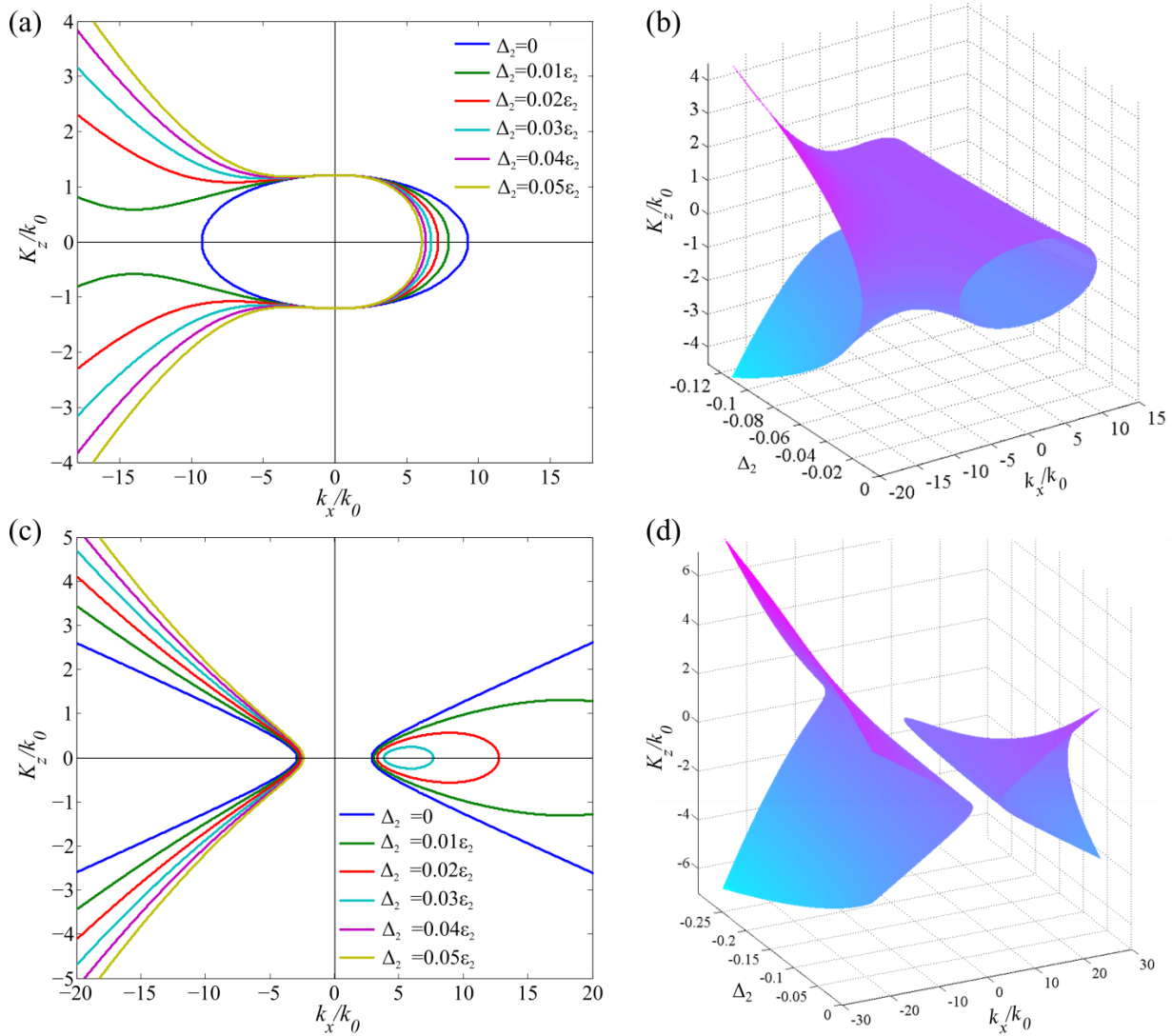

Figure 2| Changes in the isofrequency contours of nonreciprocal hyperbolic metamaterial as the result of magnetization for the wavelength of (a) $\lambda$ =455 nm and (c) $\lambda$ =360 nm. The corresponding isofrequency surfaces plotted in $k_x, K_z, \Delta$ space for (b) $\lambda$ =455 nm and (b) $\lambda$ =360 nm. The structure parameters are the same as in Fig. 1. The effect of loss is not considered for the moment.

Next, we study the effect of magnetization on the metamaterial at the wavelength $\lambda = 360$ nm, which in the absence of the external magnetization corresponds to the Type-I hyperbolic regime $\{\tilde{\epsilon}_\parallel > 0, \tilde{\epsilon}_\perp < 0\}$. Figure 2(c) shows changes in the constant frequency contours as the off-diagonal components of the metal's permittivity increases. As in the previous case, the contours acquire progressively asymmetric shapes. In particular, the left-side of the hyperbola, corresponding to the backward propagation ($k_x < 0$), opens wider as the DC magnetic field increases, and maintains its original topology. The right side, corresponding to the forward propagation ($k_x > 0$), in contrast, experiences a topological transition from the open hyperbola to the closed ellipse, which eventually collapses to a point when the magneto-optical parameter reaches a critical value of $\Delta_{cr} \equiv \Delta_2 \approx 0.04\epsilon_2$. Such magnetization induced topological transition is also illustrated by Fig.2(d), when one side of the hyperbolic contour $k_x > 0$ gradually changes to the isolated ellipse, which subsequently collapses to a single point leading to *the one-way hyperbolic regime*.

Thus we can identify three distinct nonreciprocal hyperbolic regimes. The first regime – *the nonreciprocal two-way hyperbolic regime* – is characterized by the contours consisting of two asymmetric hyperbolas (for both $k_x < 0$ and $k_x > 0$), of the topology equivalent to that of nonmagnetic Type-I structure [Fig.2(c)]. The second regime – *the forward elliptical/backward one-way hyperbolic regime* – is characterized by the hyperbola for $k_x < 0$ and the ellipse for $k_x > 0$. This regime can be subdivided into two subclasses corresponding to Type-I or Type-II hyperbolic regimes. In the case of Type-I hyperbolic regime, in addition to the hyperbolic branch, we encounter a closed and isolated ellipse of asymmetric shape [Fig.2(c)], whereas for Type-II it is half of the ellipse connected to the hyperbola [Fig.2(a)]. The third regime – *the complete one-way hyperbolic regime* – appears with the further increase of the magnetization for the Type-I hyperbolic metamaterial [Fig.2(c)] when the ellipse corresponding to $k_x > 0$ collapses to a point and then disappears above some critical value of the magneto-optical activity $\Delta > \Delta_{cr}$.

The hyperbolic dispersion originates from the coupling of surface plasmon polaritons (SPPs) supported by individual plasmonic layers comprising HMMs. Nonreciprocal hyperbolic regimes described here have the same origin, with the difference that the coupling takes place between surface magneto-plasmons, i.e. surface plasmons whose dispersion is modified by the external magnetic field. To illustrate the origin of nonreciprocal and one-way hyperbolic dispersion predicted by the effective medium theory Eqs.(1-2), and to make these predictions closer to realistic structures, we will now examine the eigenmodes and the transmission spectra calculated with the exact transfer matrix technique (described in Supplement A) for the system of a finite number of layers.

Natural plasmonic modes of layered structures are known to manifest as poles in the transmission spectra and can be clearly seen in Figs.3(a-b). Figure 3a shows the case of a single metal film (one unit cell of the HMM under study) where such poles form two continuous (low-frequency and high-frequency) dispersion curves corresponding to two magneto-plasmons predominantly residing on the opposite interfaces of the film. These modes are separated by a frequency gap originating from the asymmetric cladding of the metal layer ($\epsilon_1 \neq \epsilon_3$). Note that similar gap can also be found between so called "short-range" and "long-range" plasmons in structures with the inversion symmetric unit cell. However, in the latter case the gap originates from the coupling of the plasmons on two opposite interfaces. In the case of the asymmetric cladding considered here such coupling between plasmons still exists, but is significantly suppresses due to the mismatch in their eigenfrequencies.

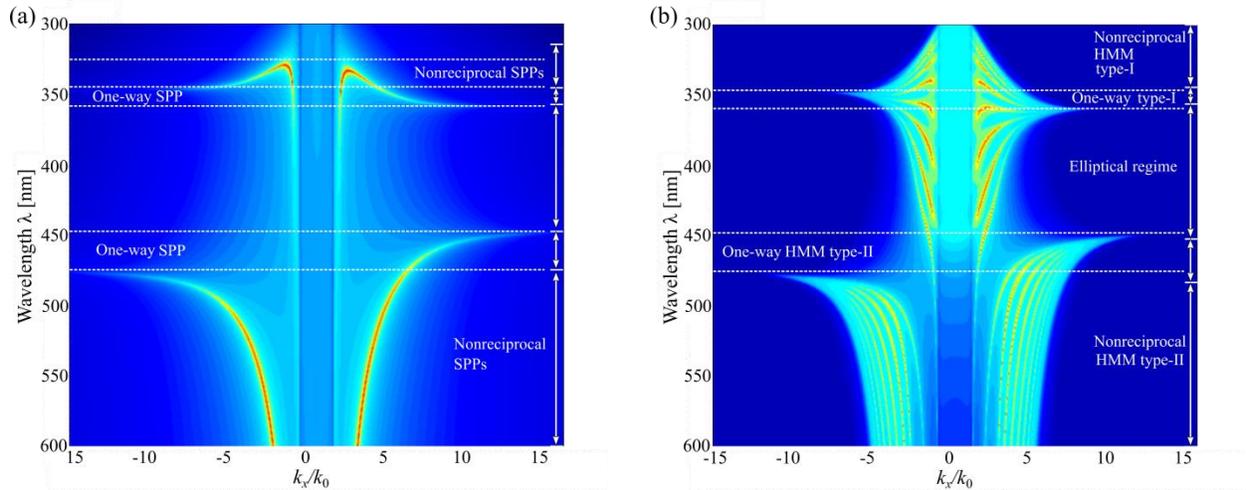

Figure 3| Poles of the transmission through (a) single unit cell (one metal layer) and (b) 10-unit cells of the HMM showing the dispersion of the magneto-plasmonic eigenmodes and indicating distinct nonreciprocal regimes. The geometry and material parameters are the same as in Fig.1.

More importantly, the dispersion of the magneto-plasmons exhibits nonreciprocity as $\lambda(\mathbf{k}) \neq \lambda(-\mathbf{k})$, and in contrast to non-magnetic case, *the curves approach different asymptotes* (indicated by dashed horizontal lines) for the forward ($k_x > 0$) and backward ($k_x < 0$) propagating waves. As it is shown below, this nonreciprocity of the magneto-plasmons of the individual unit cell is the source of the nonreciprocal hyperbolic regimes found in multilayered structures.

As the next step, we consider a structure consisting of 10 unit cells with its magneto-plasmonic bands plotted in Fig.3(b). As expected, the number of modes increases and they push each other to the domains of longer wavenumbers and shorter wavelengths, indicating onset of the hyperbolic regimes. Using conventional notations for the hyperbolic regimes, one can immediately find correspondence between the results of the effective medium outlined above and the exact calculations presented in Fig.3(b). Thus, the Type-I (Type-II) hyperbolic regime occurs due to hybridization of the high frequency (low frequency) surface magneto-plasmons of the individual metal layers. The short-wavelength Type-I and long-wavelengths Type-II regimes appear to be separated by the region of elliptical dispersion. In addition, in agreement with the effective medium calculations, the modes appear to be strongly nonreciprocal. It's interesting to note that despite the hybridization of the magneto-plasmons of individual metal layers all of the modes of multilayered structure continue approaching the same asymptotes as in the case of the single unit cell. As a result, the corresponding plasmonic bands have different "cut-off" wavelengths for the opposite propagation directions thus explaining the origin of one-way hyperbolic regimes predicted by the effective medium theory. In particular, there is a frequency window, from $\lambda \approx 450$ nm to $\lambda \approx 480$ nm, where Type-II one-way hyperbolic regime is realized, and another window, from $\lambda \approx 350$ nm to $\lambda \approx 365$ nm, where Type-I one-way regime occurs.

The hyperbolic dispersion enables many fascinating applications, from the subwavelength resolution to enhanced lasing efficiencies, which are all enabled by the presence of plasmonic modes with very long wavenumbers. Another advantage of hyperbolic metamaterials is the non-resonant origin of their unique response resulting in the broadband character of the hyperbolic regime. This makes the hyperbolic metamaterials especially promising for various applications where broadband characteristics are required. Similar arguments can apparently be applied to nonreciprocal photonic devices. Indeed, the broadband nonreciprocal response can be achieved only in bulky optical components. To our best knowledge, all attempts to reduce the footprint of nonreciprocal devices to make them more compatible with the contemporary integrated photonic components have so far relied on the use of resonant effects [18-26]. While resonances do allow enhancement of the nonreciprocal response, they also significantly reduce the operational bandwidth of the devices. Here we show that the nonreciprocal hyperbolic metamaterials do not have this limitation and may offer nonreciprocal response over a broad operational bandwidth.

Figure 4a shows transmission of the electromagnetic wave through the metamaterial (consisting of 10 unit cells) in the Type-II hyperbolic regime for the cases of forward (red line) and backward (blue line) propagation. The nonreciprocal transmission indeed occurs in the broad spectral window defined by the offset of the forward and backward transmission bands. The bandwidth of one-way response is defined by the difference in the cut-off frequencies for the forward and backward hyperbolic transmission bands shown in Fig.3, which, in its turn, is defined by the strength of magnetization. Thus, the bandwidth of one-way nonreciprocal response in the hyperbolic metamaterial is only limited by the strength of magnetic field and magneto-optical response of the materials constituting the structure, which is in sharp contrast with the bandwidth-limited nonreciprocal devices based on the resonant magneto-optical structures.

Indeed, the optical isolation and one-way response in resonant structures rely on the narrow bandwidth $\Gamma$ of a resonance whose frequency is split due to the magneto-optical activity by $\Delta\omega$ for forward and backward propagation directions so that the condition $\Delta\omega > \Gamma$ is satisfied [16]. As a result, non-uniform Lorentzian-shaped one-way transmission occurs over $\Gamma$-wide band. As opposed to this principle of operation, the nonreciprocal HMM device provides a uniform transmission over the entire frequency range $\Delta\omega$ [Fig. 4(a)]. Taking into account the fact that with a sufficient number of layers the hyperbolic transmission bands can be made arbitrarily wide, one can always design a nonreciprocal device with a desirable bandwidth, provided the magneto-optical response of a sufficient strength is available.

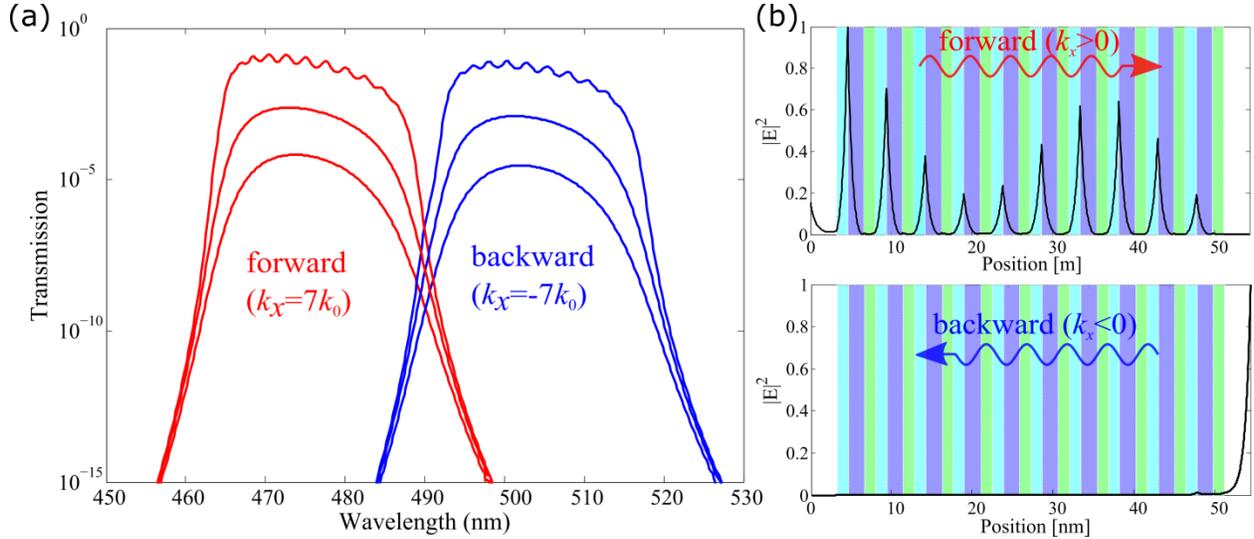

Figure 4| (a) Transmission spectrum and (b) the field distribution at $\lambda$=470 nm for the forward ($k_x = 7k_0$) and backward ($k_x = -7k_0$) propagation directions calculated by the transfer matrix technique for the HMM consisting of 10 unit cells. The material parameters used are the same as in Fig.1, and in subplot (a) the damping frequency is changed from $\gamma$ =1.13×10$^{13}$ [rad/s] to $\gamma$ =2.83×10$^{13}$ [rad/s] to $\gamma$ =5.65×10$^{13}$ [rad/s].

As for any other plasmonic structure, ohmic losses will play a detrimental role for operation of the nonreciprocal hyperbolic metamaterial. For example, Fig. 4(a) shows how the transmission through the metamaterial changes with the increase of the damping frequency, and, while the bandwidth of the nonreciprocal response stays nearly unchanged, the transmission drops significantly. To avoid this decrease in the transmission, the number of layers in the HMM should be reduced, which, on the other hand, will narrow the bandwidth of the hyperbolic transmission band. Therefore, in addition to the magnetic response strength, the limitation in the operational bandwidth of nonreciprocal HMM devices will be also dictated by the losses. Nevertheless, it's apparent that the operational bandwidth of HMM can always be made superior as compared to that in resonant plasmonic structures where losses have similar or worse detrimental effects.

To understand the origin of the one-way transmission through the hyperbolic bands it's instructive to inspect the electric field distribution inside the nonreciprocal HMM. For example, the field distribution inside the structure, calculated for the wavelength $\lambda = 470$ nm, is plotted in Fig. 4(b) for forward ($k_x = 7k_0$) and backward ($k_x = -7k_0$) propagation directions. As expected, the transmission of the evanescent field in the forward direction occurs due to the excitation of the side-coupled surface plasmons (on the left side of metallic layers) propagating along the layers in the upward direction ($k_x > 0$), which transfer the electromagnetic energy through the structure from left to right. On the other hand, when excited from the opposite side with the backward propagation direction ($k_x < 0$), the excitation wavelength happens to exceed the cut-off wavelength ($\lambda_0 =480$ nm) and no plasmonic modes are excited, resulting in the fast decay of the field inside the structure and vanishingly small transmission. These field profiles suggest another class of application which can be possible even for strongly absorbing structures, such as nonreciprocal and one-way absorbers. Thus, according to Fig.4(b) the magneto-plasmons in the metamaterial are excited only for forward but not backward excitation which implies that the absorption of incident evanescent field will take place only in the former case, while in the latter, strong reflection will occur.

In summary, we demonstrated theoretically for the first time the possibility of nonreciprocal light transmission using magnetoplasmonic hyperbolic metamaterials. New nonreciprocal hyperbolic regimes and one-way topological transitions between hyperbolic and elliptical dispersion regimes were found. Due to the non-resonant nature of the metamaterial, a uniform broadband character of the transmission was achieved in the one-way hyperbolic regime, which envisions a significant potential for practical applications. In addition to the visible domain studied here, the results presented hold a great promise for applications at the infrared and terahertz frequencies where nonreciprocal hyperbolic metamaterials are made of highly doped semiconductors.

**References**

[1] J. B. Pendry, Negative refraction makes a perfect lens. Phys. Rev. Lett. **85**, 3966 (2000).

[2] D.R. Smith, J.B. Pendry, & M.C.K. Wiltshire, "Metamaterials and Negative Refractive Index", Science **305**, 788–792 (2004).

[3] V. M. Shalaev, Optical negative-index metamaterials. Nat. Photonics **1**, 41 (2007).

[4] Engheta, N. & R.W. Ziolkowski, "Metamaterials: physics and engineering explorations", Wiley & Sons (New Jersey) 2006.

[5] A.K. Sarychev & V.M. Shalaev, "Electrodynamics of Metamaterials", World Scientific, Singapore 2007.

[6] W. Cai & V. Shalaev, "Optical Metamaterials: Fundamentals and Applications", Springer, Berlin 2009.

[7] R. J. Potton, "Reciprocity in optics", Rep. Prog. Phys. **67**, 717–754 (2004).

[8] D. Jalas, A. Petrov, M. Eich, et al., "What is - and what is not - an optical isolator", Nat. Photon. **7**, 579–582 (2013).

[9] H. A. Haus, "Waves and Fields in Optoelectronics", Prentice-Hall, 1984.

[10] B. E. A. Saleh & M. C. Teich, "Fundamentals of Photonics", Wiley-Interscience, 2007.

[11] L. D. Tzuang, K. Fang, P. Nussenzveig, S. Fan, & M. Lipson Nat. Photon. **8**, 701–705 (2014).

[12] M. Inoue, K. Arai, T. Fujii, & M. Abe, "One-dimensional magnetophotonic crystal", J. Appl. Phys. **85**, 5768-5770 (1999).

[13] A. Figotin & I. Vitebskiy, "Nonreciprocal magnetic photonic crystals", Phys. Rev. E **63**, 06660901, (2001).

[14] Z. Wang & S. Fan, "Optical circulators in two-dimensional magneto-optical photonic crystals", Opt. Lett. **30**, 1989–1991 (2005).

[15] Z. Yu, Z. Wang, & S. Fan, "One-way total reflection with one-dimensional magneto-optical photonic crystals", Appl. Phys. Lett. **94**, 171116:1-3 (2009).

[16] A. B. Khanikaev & M. J. Steel, "Low-symmetry magnetic photonic crystals for nonreciprocal and unidirectional devices", Opt. Exp. **17**, 5265-5272 (2009).

[17] A.B. Khanikaev, A.V. Baryshev, M. Inoue, & Yu. S. Kivshar', "One-way electromagnetic Tamm states in magnetophotonic structures", Appl. Phys. Lett. **95**, 011101:1–3 (2009).

[18] V. I. Belotelov, L. L. Doskolovich, & A. K. Zvezdin, "Extraordinary Magneto-Optical Effects and Transmission through Metal-Dielectric Plasmonic Systems", Phys. Rev. Lett. **98**, 077401:1–4 (2007).

[19] A. B. Khanikaev, A. V. Baryshev, A. A. Fedyanin, A. B. Granovsky, & M. Inoue, "Anomalous Faraday effect of a system with extraordinary optical transmittance", Opt. Exp. **15**, 6612–6622 (2007).

[20] Z. Yu, G. Veronis, Z. Wang, & S. Fan, "One-Way Electromagnetic Waveguide Formed at the Interface between a Plasmonic Metal under a Static Magnetic Field and a Photonic Crystal", Phys. Rev. Lett. **100**, 023902:1–4 (2008).

[21] G. A. Wurtz, W. Hendren, R. Pollard, et al., "Controlling optical transmission through magneto-plasmonic crystals with an external magnetic field", New J. Phys. **10**, 105012:1–10 (2008).

[22] A. B. Khanikaev, H. Moussavi, G. Shvets, & Yu. S. Kivshar', "One-way extraordinary optical transmission and nonreciprocal spoof plasmons", Phys. Rev. Lett. **105**, 126804:1–4 (2010).

[23] A. Davoyan & N. Engheta, "Nonreciprocal Rotating Power Flow within Plasmonic Nanostructures, " Phys. Rev. Lett. **111**, 047401 (2013).

[24] A.R. Davoyan & N. Engheta. "Nanoscale plasmonic circulator", New J. Phys. **15**, 083054 (2013).

[25] J.Y. Chin, T. Steinle, T. Wehlus, et al., "Nonreciprocal plasmonics enables giant enhancement of thin-film Faraday rotation", Nat. Comm. (2013). DOI:10.1038/ncomms2609

[26] U. K. Chettiar, A.R. Davoyan, & N. Engheta, "Hotspots from nonreciprocal surface waves", Optics Letters **39**, 1760-1763 (2014).

[27] S. H. Mousavi, A. B. Khanikaev, J. Allen, M. Allen, & G. Shvets, "Gyromagnetically induced transparency of Metasurfaces", Phys. Rev. Lett. **112**, 117402:1-5 (2014).

[28] H. Zhao, J. Zhou, Q. Zhao, et al., "Magnetotunable left-handed material consisting of yttrium iron garnet slab and metallic wires", Appl. Phys. Lett. **91**, 131107 (2007).

[29] F. J. Rachford, D. N. Armstead V. G. Harris, & C. Vittoria, "Simulations of Ferrite-Dielectric-Wire Composite Negative Index Materials", Phys. Rev. Lett. **99**, 057202 (2007).

[30] L. Kang, Q. Zhao, H. Zhao, & J. Zhou, "Magnetically tunable negative permeability metamaterial composed by split ring resonators and ferrite rods", Opt. Express **16**, 8825 (2008).

[31] W. Li, Z. Liu, X. Zhang, & X. Jiang, "Switchable hyperbolic metamaterials with magnetic control", Appl. Phys. Lett. **100**, 161108 (2012).

[32] K. Bi, J. Zhou, H. Zhao, X. Liu, & C. Lan, "Tunable dual-band negative refractive index in ferrite-based metamaterials", Opt. Express **21**(9), 10746-10752 (2013).

[33] Y. G. Huang, G. Wen, W. Zhu, et al., "Experimental demonstration of a magnetically tunable ferrite based metamaterial absorber", Opt. Express **22**(13), 16408-17 (2014).

[34] Z. Jacob, L. V. Alekseyev, & E. Narimanov, "Optical Hyperlens: Far-field imaging beyond the diffraction limit", Opt. Express **14**, 8247 (2006).

[35] Z. Liu, H. Lee, Y. Xiong, C. Sun, & X. Zhang, "Far-field optical hyperlens magnifying sub-diffraction-limited objects", Science **315**, 1686 (2007).

[36] A. J. Hoffman, L. Alekseyev, S. S. Howard, et al., "Negative refraction in semiconductor metamaterials", Nat. Mater. **6**, 946 (2007).

[37] M. A. Noginov, H. Li, Yu. A. Barnakov, et al., "Controlling spontaneous emission with metamaterials", Opt. Lett. **35**, 1863 (2010).

[38] Z. Jacob, J Y. Kim, G V. Naik, et al., "Engineering photonic density of states using metamaterials", Appl. Phys. B **100**, 215 (2010).

[39] A. N. Poddubny, P. A. Belov, & Y. S. Kivshar, "Spontaneous radiation of a finite-size dipole emitter in hyperbolic media", Phys. Rev. A **84**, 023807 (2011).

[40] O. Kidwai, S. V. Zhukovsky, & J. E. Sipe, "Dipole radiation near hyperbolic metamaterials: Applicability of effective-medium approximation", Opt. Lett. **36**, 2530 (2011).

[41] V. Drachev, V. A. Podolskiy, & A. V. Kildishev, "Hyperbolic Metamaterials: new physics behind a classical problem," Opt. Express **21**(12), 15048–15064 (2013).

[42] C. L. Cortes, W. Newman, S. Molesky, & Z. Jacob, "Quantum nanophotonics using hyperbolic metamaterials", J. Opt. **14**, 063001:1-15 (2012).

[43] H. N. S. Krishnamoorthy, Z. Jacob, E. Narimanov, I. Kretzschmar, & V. M. Menon, "Topological Transitions in Metamaterials", Science **336**, 205-209 (2012).

[44] A. Poddubny, I. Iorsh, P. Belov, & Y. Kivshar, "Hyperbolic metamaterials", Nat. Photon. **7**, 948–957 (2013).

[45] J. Elser, V. A. Podolskiy, I. Salakhutdinov, & I. Avrutsky, "Nonlocal effects in effective-medium response of nanolayered metamaterials", Appl. Phys. Lett. **90**, 191109 (2007).